\documentclass[12pt]{iopart}
\usepackage{graphicx}

\usepackage{iopams}  
\def\etal{{\it et\thinspace al.}\ }
\def\eion{{(e~+~ion)}\ }

\def\fexvii{{\rm Fe~\sc xvii}\ }
\def\fexviii{{\rm Fe~\sc xviii}\ }
\def\fexix{{\rm Fe~\sc xix}\ }

\newcommand{\be}{\begin{equation}}
\newcommand{\ee}{\end{equation}}
\begin{document}

\title[R-Matrix calculations for opacities~IV]{R-Matrix calculations for
opacities:~IV.: Convergence, completeness, and comparison 
of relativistic R-matrix and distorted wave calculations for 
\fexvii and \fexviii} 

\author{L Zhao$^1$, S N Nahar$^1$, A K Pradhan$^{1,2,3}$}

\address{$^1$ Department of Astronomy,
Ohio State University, Columbus, Ohio 43210, USA}
\address{$^2$ Biophysics Graduate Program, $^3$ Chemical Physics
Program, The Ohio State University, Columbus, Ohio 43210, USA}
\vspace{10pt}

\begin{abstract}
        To investigate the completeness of coupled channel (CC) 
Breit-Pauli R-Matrix
(BPRM) calculations for opacities, 
we employ the relativistic distorted wave (RDW) method 
to complement (``top-up'') and compare the BPRM photoionization cross sections 
for high-$n\ell$ levels of both  \fexvii and \fexviii. Good
agreement is found in background photoionization cross sections using these two
methods, which also ensures correct matching of bound level cross
sections for completeness. In order to top-up
the CC-BPRM calculations, bound-bound transitions involving
additional bound levels, and a large number of 
doubly-excited quasi-bound levels corresponding to 
BPRM autoionizing resonances described in paper RMOPII, 
are calculated using
the RDW method. Photoionization cross sections in the high energy region are
also computed and compared up
to about 500 $Ry$, and contributions from higher core level
excitations than BPRM are considered.
The effect of configuration interaction is investigated,
which plays a significant role in
correctly reproducing some background cross sections.
Owing to the fact that the additional RDW levels correspond to
high-$n\ell$ bound levels that are negligibly populated according to the
Mihalas-Hummer-D\"{a}ppen equation-of-state (paper I), the effect on
opacities is expected to be small.
\end{abstract}

%
%
%
%
%

\section{Introduction}

Previous papers I-III in this series (hereafter RMPP1, RMOP2, RMOP3) reported
BPRM calculations and plasma effects related to iron opacity 
at conditions similar to the solar 
radiation/convection zone boundary or the base of the convection zone
(BCZ). As outlined in paper RMOP1, opacity calculations need to consider 
all possible
mechanisms for photon absorption and scattering from all atomic
constituents, including all levels that might possibly contribute. 
Furthermore, in order to resolve discrepancies among various theoretical
models based on the DW methods that include different sets of transition
arrays, and experimental measurements \cite{p18,b15}, it is
necessary to establish convergence of the BPRM calculations 
and completeness of transitions considered.

Extensive CC-BPRM calculations
$R$-Matrix (BPRM) calculations were carried out for \fexvii
including 60 fine-structure levels within the $n\leq 3$ complexes in the
\fexviii target ion \cite{n11}, and 99 $LS$
terms within $n\leq 4$ (99LS-RM). They show strong photon absorption due
to core excitation, resulting in an increment of 35\% in the Rosseland
mean opacity over the Opacity Project (OP) data \cite{np16a}. 
 Whereas these previous calculations demonstrated that in the
$R$-Matrix opacity calculations convergence of the close-coupling
expansion was a necessary condition for accuracy, completeness of all
possible excited configurations by
additional contributions in the high-energy region still remains to be
ascertained \cite{np16a,np16b,ih17}.
At BCZ conditions \fexvii, \fexviii and \fexix 
are the three dominant iron ions. For example, at 
the measured iron opacity \cite{b15} and temperature and density 
T=$2.1\times10^6$ and N$_e$=$3.1\times10^22$cc, the three 
ionization fractions are 0.19, 0.38
and 0.29, respectively \cite{np16a}. 

  In this paper, we consider the 218CC-BPRM calculation for \fexvii and 
276CC-BPRM calculation for
\fexviii, as described in paper RMOP2. The additional or topup
transitions for bound-bound and bound-free data are
obtained from relativistic distorted wave (RDW) calculations 
using the flexible atomic code (FAC) \cite{gu08}.
In the following sections, the specifications of the 218CC- and 276CC-BPRM 
calculations of paper RMOP2 are summarized, followed by the top-up 
configurations and transitions calculated using FAC. To ensure data 
correspondence from FAC, a procedure of matching the bound levels from 
BPRM and FAC results is described, and the bound-bound and bound-free top-up 
calculations detailed afterwards. 

A key step in the matching-topup procedure is level identification.
Unlike atomic structure and DW calculations, BPRM calculations do not
assign spectroscopic designations {\it a priori} and bound states are
obtained only as eigenvalues of the \eion Hamiltonian. As described in
paper RMOP2, the code BPID (Fig.~1, RMOP-I) is used to obtain relevant
parameters and for spectroscopic identification of levels computed in
BPRM calculations. The RDW calculations of course have a pre-assigned
identification based on initial set of electronic configurations
specified in the configuration-interaction (CI) basis. In some instances
exited level configuration mixing is such that one
configuration does not dominate the wavefunction expansion of a given
state and RDW and BPRM assignations do not match. It is then required to
carefully examine level parameters such as quantum defects 
and associated bound-bound and bound-free transitions to ascertain
matching data. 
 Another consideration is that the precise number of BPRM bound-state
eigenvalues depends on an energy mesh or effective quantum number
$\nu(E)$ obtained by "scanning" at a fine mesh with sufficient refinement
to ensure convergence. The procedure and results are discussed in this
paper.

A potentially important factor is that
the close coupling approximation introduces autoionizing resonances in
photoionization cross section, which may be affected by radiative
damping in highly charged H- or He-like ions, and thereby reduce
effective cross sections considerably \cite{pz97}. However, radiation
damping occurs {\it after} photoabsorption, and for ions such as
\fexvii and \fexviii this effect is negligible \cite{z98, znp99}. 
Therefore, undamped
photoionization cross sections are used in opacity calculations, as
reported in paper RMOP2. 

\section{BPRM bound-free and bound-bound data} \label{sec:bprm}

The current BPRM calculations for \fexvii and \fexviii are unprecedented
in terms of scope and magnitude of data produced and processed for
opacity calculations, with the maximum number of free channels 998 and 1288 
respectively, from calculations reported in RMOP2. For \fexvii,
99LS-RM calculation \cite{np16a} is extended to 218CC-BPRM by including
the fine structure of the target states. The target configurations ($1s$
is always full, so omitted for brevity) included are ${2s^2 2p^5}$, ${2s
2p^6}$, ${2s^2 2p^4 n\ell}$, ${2s 2p^5 n\ell}$, ${2p^6 3\ell'}$, where
$n=3, ~4$, and $\ell,~ \ell'\leq2$, which have 99 $LS$ terms, or 218
fine structure levels. The continuum orbitals included are $\ell\leq 9$,
and the number of continuum $R$-Matrix basis functions included is 20.
The bound states are found by scanning the eigenvalues of the \eion
Hamiltonian on an effective quantum number $\nu$
up to $\nu \leq 10.1$ \cite{s85}. However, as mentioned above,
unlike atomic structure calculations where electronic configurations
are specified {\it a priori}, R-matrix calculations do not 
provide spectroscopic spin-orbital quantum number designations for the bound
levels obtained, nor guarantee that all possible bound levels are found
within the $\nu$-range of interest. 

To resolve the first issue, the 
computer program BPID \cite{np00} 
has been developed as part of the RM opacity codes
described in paper P1. Using the code BPID
one can identify most of the bound levels spectroscopically, albeit with a few
remaining highly mixed levels undetermined (viz. \cite{n00}). 
That obstacle might be 
overcome by comparing some physical quantities of these levels calculated
by an atomic structure code such as SUPERSTRUCTURE \cite{ejn}
and FAC \cite{gu08}, 
as for example for photoionization cross section to be described in the 
next section. The second issue depends on the scanning $\Delta-\nu$-mesh 
employed; $\Delta \nu=0.001$ yields 
fewer bound levels in the 218CC-BPRM than 60CC-BPRM, so a finer step 
0.0001 or 0.0005 is used in the region where levels are missing, 
which finally gives 464 bound levels, 10 more than 60CC-BPRM. 

For the larger \fexviii case, 
two sets of BPRM calculations are done with different
target configurations. One includes up to $n=3$ target configurations,
i.e. $2s^2 2p^4$, $2s 2p^5$, $2p^6$, $2s^2 2p^3 3\ell$, $2s 2p^4 3\ell$,
where $\ell\leq2$, which yields 200 target fine structure levels. In
addition to target configurations above, the other BPRM calculation
includes $n=4$ configurations, i.e. $2s^2 2p^3 4\ell$, where
$\ell\leq2$, which yield 276 fine structure levels. The parameters set
for the continuum orbitals and basis functions are the same as for
\fexvii 218CC-BPRM calculation. The 200CC-BPRM calculation finds 
1149 bound levels with $\Delta \nu = 0.001$, while 276CC-BPRM calculation 
finds 1163 bound levels with 0.001 as the initial attempt in $\nu$-mesh,
and 0.0001 or 0.0005 as the second attempt, 
in the region where levels are missing compared with 200CC-BPRM. Thus we
may be confident of having converged with respect to possible number of
bound levels with BPRM calculations.

\begin{figure}
\vspace{0.0in}
\includegraphics[width=0.50\linewidth,height=7cm]{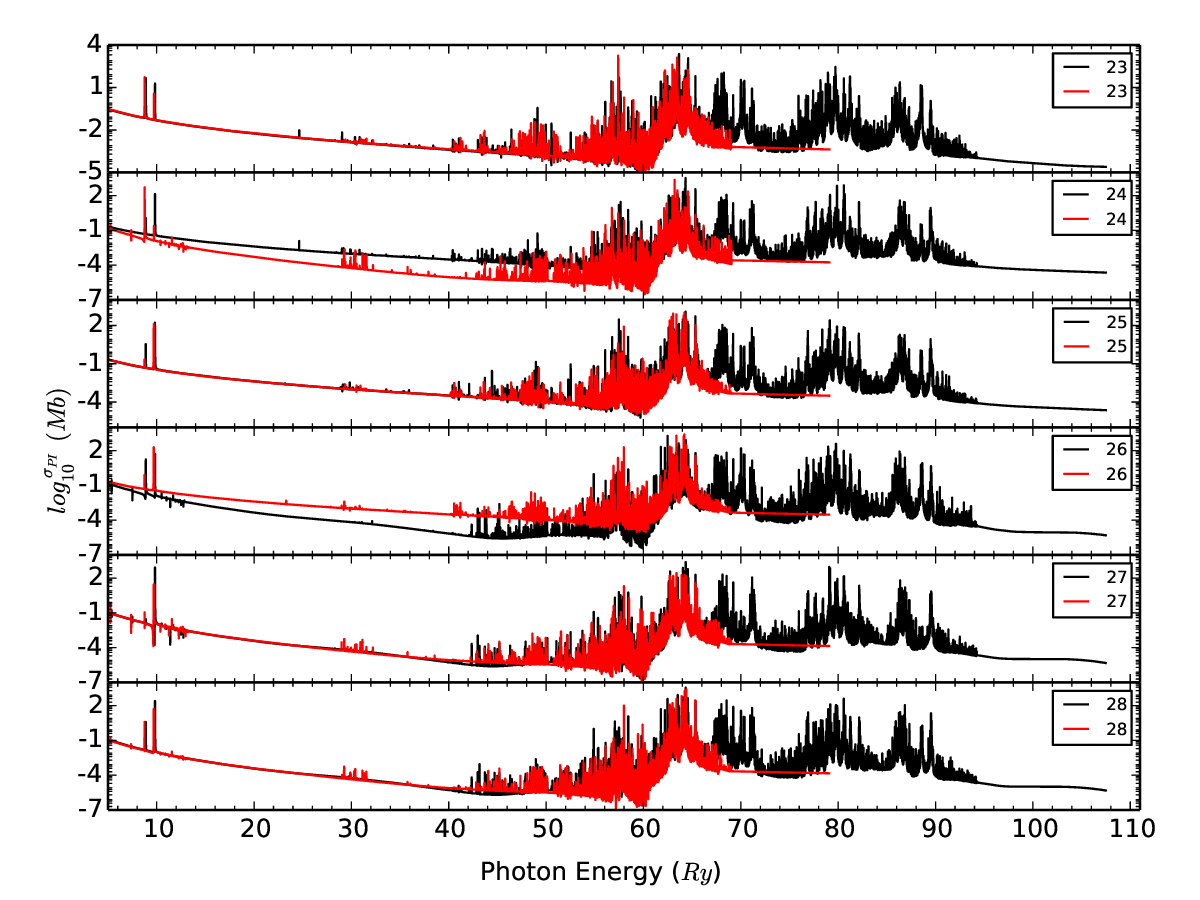}
\includegraphics[width=0.50\linewidth,height=7cm]{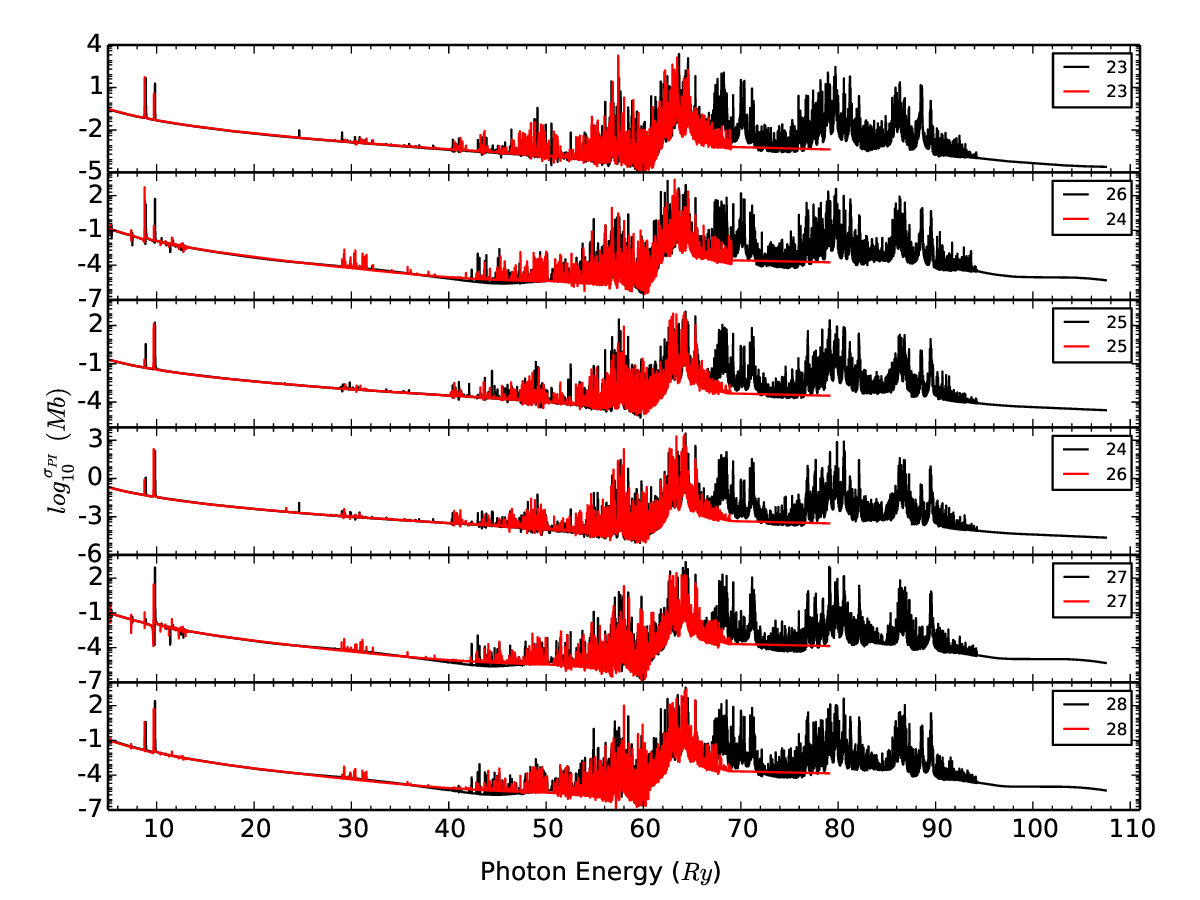}
\includegraphics[width=0.50\linewidth,height=7cm]{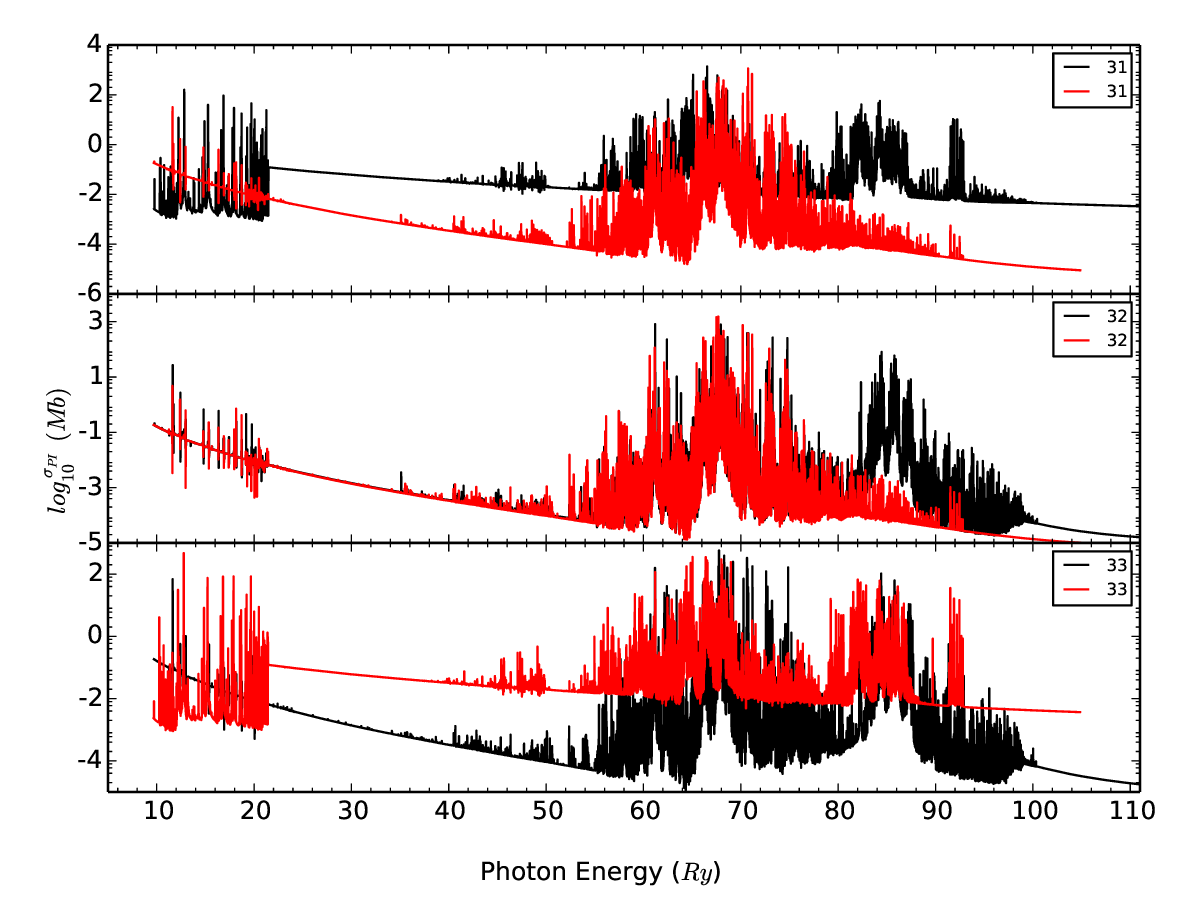}
\includegraphics[width=0.50\linewidth,height=7cm]{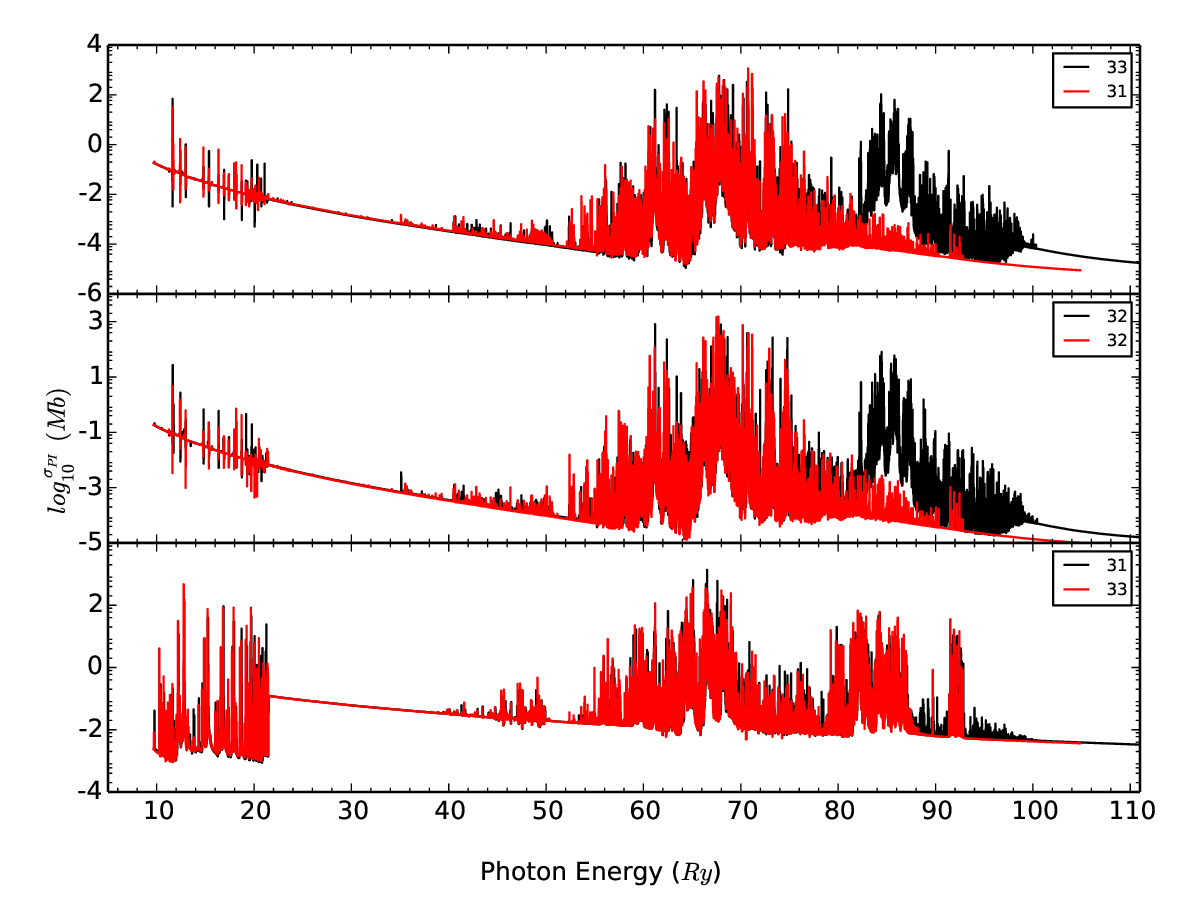}
	\caption{Photoionization cross section of 6 closely packed levels for 
\fexvii and 3 levels for \fexviii before and after being switched. 
60CC-BPRM (red), 218CC-BPRM (black); 200CC-BPRM (red), 276CC-BPRM
(black) \label{fig:rematch}}
\end{figure}

To compute iron opacities for \fexvii and \fexviii oscillator strengths 
from the 60CC-BPRM
and 200CC-BPRM calculations, and photoionization cross sections from
the 218CC-BPRM and  276CC-BPRM calculations, are used respectively
(paper RMOP2). So
in doing the FAC top-up calculations, matching the bound levels for each
BPRM calculation is necessary but complicated by the fact that 
they have different number of
bound levels, especially energy regions where levels are densely packed
(see table \ref{tab:packedlevels}) and the order of their spectroscopic
designations may be mismatched and needs to be shuffled 
(see figure~\ref{fig:rematch}).\footnote{It bears
emphasis that the opacities {\it per se} are independent of any
spectroscopic labels; however, they are necessary for processing the
bound-bound and bound-free radiative atomic transitions, 
and for comparing with other data sources.} Photoionization cross
sections of 6 levels of \fexvii are plotted in figure~\ref{fig:rematch}
for 60CC-BPRM and 276CC-BPRM, and we find distinct difference in level 24 and
26. Similar issue arises in figure~\ref{fig:rematch} of \fexviii. 
Even though these levels have similar energy, they may have distinctive
configurations (see section \ref{sec:sectionmatching}), which is the reason
why we should redo the identification for different CC-BPRM
calculations. After being switched, these levels show good agreement (see
figure~\ref{fig:rematch}).


\begin{table}
	\centering
	\caption {Selected packed levels of \fexvii ($J = 4, \pi = 0$) and 
\fexviii ($J=5/2, \pi=0$) (Note: the energy is $z$-scaled, and in unit of $10^{-2} Ry$)}
	\begin{tabular}{|c || c | c | c|}
		\hline
		 & level index & 60CC-BPRM & 218CC-BPRM \\
		 \hline
\fexvii & 23 & -1.242945 & -1.239702 \\
													 & 24 & -1.239540 & -1.238049 \\
													 & 25 & -1.237855 & -1.236506 \\
													 & 26 & -1.236421 & -1.235733 \\
													 & 27 & -1.235081 & -1.235227 \\
													 & 28 & -1.234742 & -1.234723 \\
		\hline
		\hline
		& level index & 200CC-BPRM & 276CC-BPRM \\
		\hline
\fexviii & 31 & -3.996445 & -4.004678 \\
													 & 32 & -3.993010 & -4.003468 \\
													 & 33 & -3.989362 & -4.000058 \\
		\hline
	\end{tabular}	
	\label{tab:packedlevels}
\end{table}


\section{Complementarity between BPRM and RDW: the top-up procedure}
The Opacity Project \cite{op} employed a small \eion wavefunction
expansion including outer open Shell configurations
in the close-coupling approximation and
$R$-Matrix method in LS-coupling 
to calculate non-relativistic photoionization cross sections in the low
energy region, and adopted the Kramer approximation ("tail") 
to fit and extend to higher energies afterwards.
In previous calculations, \cite{z98} replaced the Kramer tails with the
RDW results including the contribution from inner-shell processes.
Later opacity tables were updated by also including inner-shell
transitions \cite{bs03}. In this section, we describe the procedure 
employed to compare and complement BPRM data with RDW data from FAC. 
That requires careful matching between BPRM and 
FAC cross sections for all bound levels, 
and detailed bound-bound and bound-free top-up
calculations. We also discuss the effect of configuration 
interaction on photoionization cross sections. 

\subsection{Matching} \label{sec:sectionmatching}
In BPRM calculations, bound levels with continuum orbitals $\ell\leq9$ 
and effective quantum number $\nu\leq10.1$ are formed by coupling 
the $n=2$ core states with the continuum $\nu$ and $\ell$ of the outer
electron. So in FAC we set the bound configurations as a permutation 
of the $n=2$ core configurations and an outer electron with principle 
quantum number $n\leq10$. With the same n-complex configuration 
interaction included, atomic structure is solved and sorted by total 
angular momentum $J$ and parity $\pi$, and ordered in energy, we find 
excellent agreement in the energy between BPRM and FAC values.

In calculating photoionization cross sections we include the whole
$n$-complex of core configurations for CI
(hereafter CI)
purpose, but only the transitions to core configurations that are
included in BPRM calculations. To delineate 
photoionization cross sections at the edges of energy grid, 
the energy mesh is created in such a
way that within any two adjacent thresholds 10 points are uniformly
assigned. The partial photoionization cross section is the computed in the
default 6 energy grids, and interpolated/extrapolated in our mesh, and
summed to give total cross sections for each bound
level. To investigate the effect of CI,
two sets of RDW calculation are
carried out. Both sets only allow the same-$n$-complex configuration
interaction for bound configurations, but for the core configurations
one of them only allows the same-$n$-complex CI
and the other allows different $n$-complexes.
We mix all the core configurations together. In RDW calculations, the
photoionization cross section is related to the dipole operator matrix
$<\psi_i|\mathbb{D}|\psi_f>$. The $|\psi_i>$ involves 
the electron in the bound state
to the continuum and all other electrons, and
must stay the same if only same-$n$-complex configuration
interaction is allowed. It can be different if different-$n$-complex
CI is considered \cite{gu08}.

To match bound state levels from BPRM and RDW, it is necessary to
compare cross sections to ensure the correctness of
matching. We plot BPRM and RDW photoionization cross sections in
the order of energy for each $J$, $\pi$ symmetry pair, and a level is
matched when the energy and the photoionization cross section agree
reasonably well (here the background of the BPRM data and RDW are
compared). Photoionization cross sections of majority of
bound levels show excellent consistency at the first attempt (see
figure~\ref{fig:rematch} for \fexvii and \fexviii. 
The $LS$-term notation ( $S$ and $L$) can not be
determined from FAC output for all levels, so only configuration and
total angular momentum $J$ are given). However, when several levels are
almost degenerate (see table \ref{tab:tablelevlsmatching}),
distinctive differences may yet occur in cross sections.
Such levels need to be switched till good matching is achieved
(See figure~\ref{fig:fighowtomatch} for \fexvii and \fexviii
The process is justified since level identification of
near-degenerate BPRM levels may not exactly correspond in energy from
different 
atomic structure codes such as FAC, since spectroscopic
designations depend on CI included and coupling
schemes employed for the \eion system\footnote{All BPRM photoionization
cross sections include a small region below the
lowest ionization threshold for each level \cite{op}, where no RDW
data are shown.}.

In table \ref{tab:tablelevlsmatching}, we can see the
energy levels computed in BPRM and RDW agree quite well. For
\fexvii, levels 13 and 14 , and levels 15 and 16 lie very close
to each other, and in figure~\ref{fig:fighowtomatch}, levels 13 and 16
achieve good agreement, while levels 14 and 15 don't. So we switch the
order of levels 14 and 15 in RDW calculation and recompare with
good agreement. Thus levels 13 - 16 in BPRM calculation are
matched with those in RDW calculation. The same procedure is applied to
levels 65 and 66 of \fexviii in table \ref{tab:tablelevlsmatching}, and
the result is show in figure~\ref{fig:fighowtomatch}.

\begin{figure}
        \centering
\includegraphics[width=\textwidth,height=7cm]{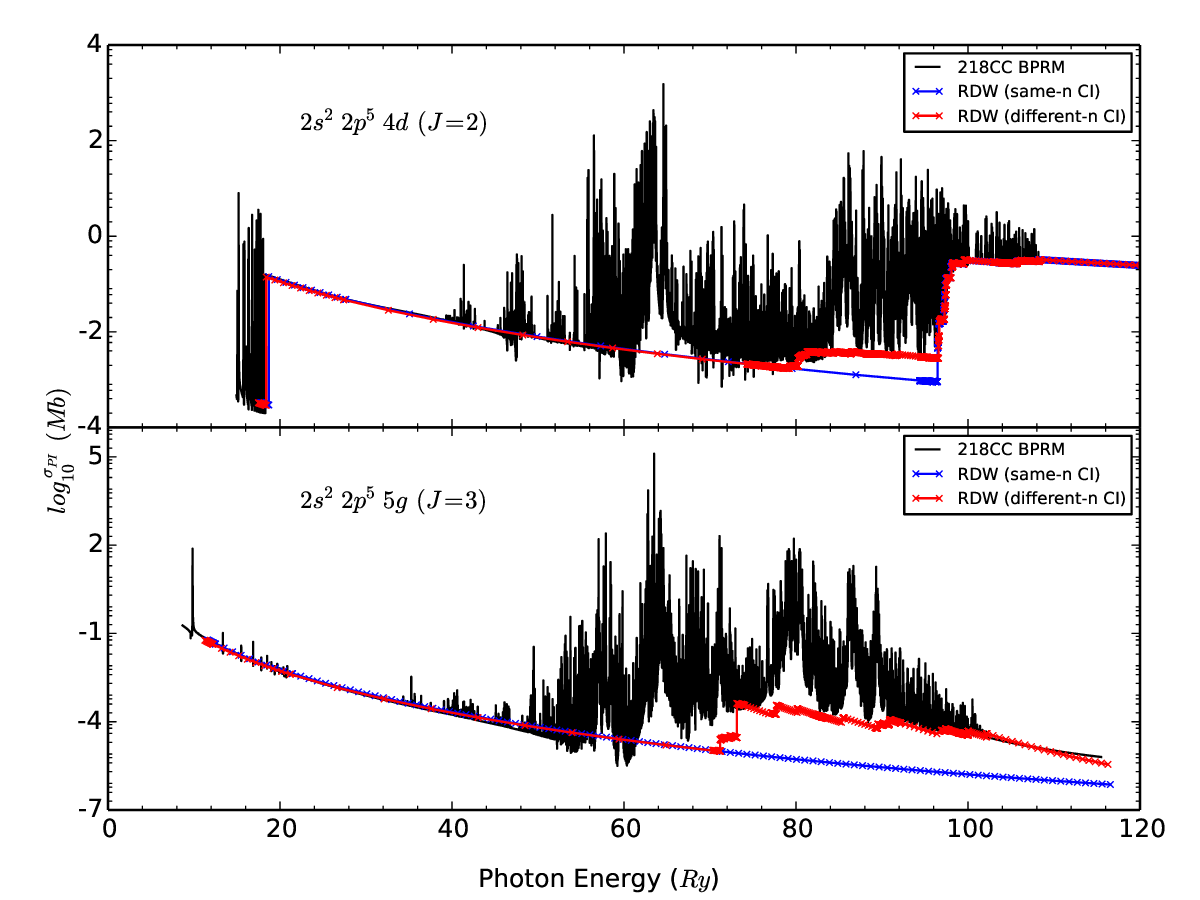}
\includegraphics[width=\textwidth,height=7cm]{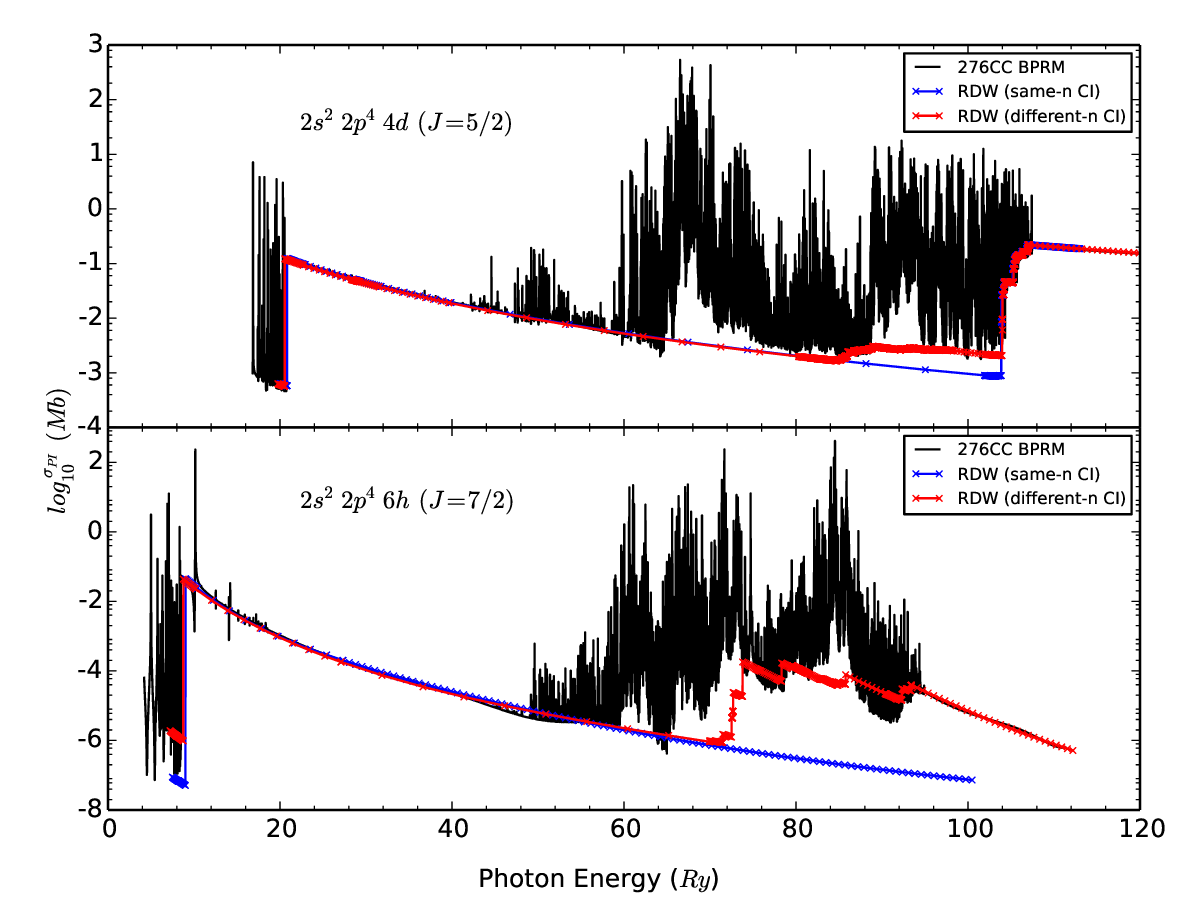}
        \caption{Most of the levels are matched at the first attempt
with excellent consistency in photoionization cross section. The
configuration is attached with each level. BPRM (black), RDW(blue and
red). ``Same-n CI'' means only same-n-complex CI
is considered for core configurations, and ``different-n CI'' refers to
both same-n- and different-n-complex CI is
considered.}
        \label{fig:figmatch}
\end{figure}

\begin{figure}
        \centering
\includegraphics[width=\textwidth,height=9cm]{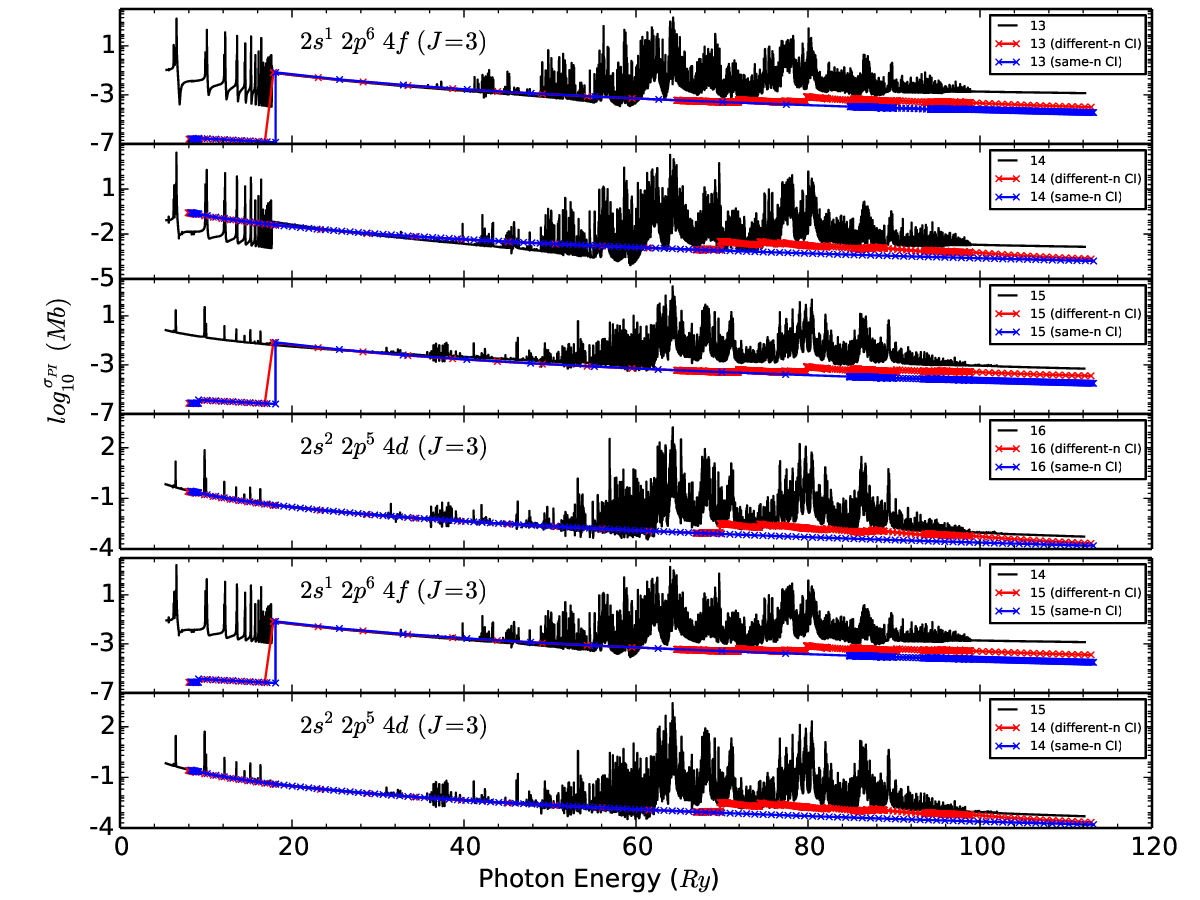}
\includegraphics[width=\textwidth,height=9cm]{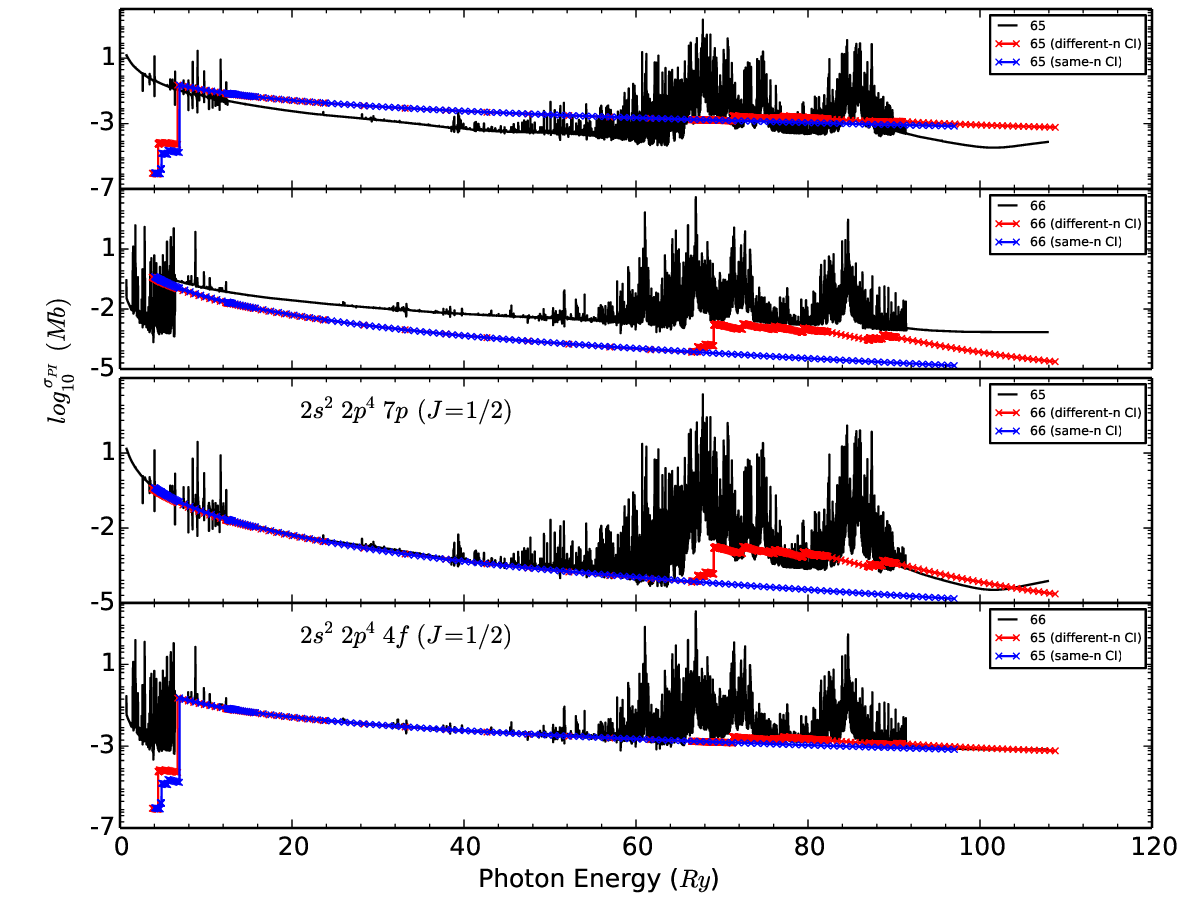}
        \caption{Multiple attempts are needed to ensure the correct
matching when the levels are found with discrepancy in the
photoionization cross section. In ?? and
?? the discrepancy is shown in upper panel and
the final matching is in the lower one. Find the configuration attached
for each level. BPRM (black), RDW(blue and red). ``Same-n CI'' refers to
only same-n-complex CI is considered for core
configurations, and ``different-n CI'' refers to both same-n- and
different-n-complex CI are considered.}
        \label{fig:fighowtomatch}
\end{figure}

In figures~\ref{fig:figmatch} and \ref{fig:fighowtomatch}, we show two sets of
RDW calculation as described above, and study the effect of configuration
interaction on photoionization cross sections, with the upper panel
of figure~\ref{fig:figmatch} as an example. The dominant configuration of the
bound state after being matched with RDW is $2s^2 2p^5 4d$, so with only
the same-$n$-complex CI of core configurations
considered, the transitions can only happen to core configurations $2s^2
2p^5$, $2s 2p^5 4\ell$, $2s^2 2p^4 4\ell$ and $2p^6 4\ell$, where
$\ell=s, p, d$, while with different-$n$-complex CI
of  core configurations additional contribution can be
from all other core configurations. From  the upper panel of 
figure~\ref{fig:figmatch} we can see that the same-$n$-complex configuration
interaction gives reasonably good background, though with some big gaps,
while different-$n$-complex CI fills up the big
gap and improves the background significantly. Similar phenomenon
can be found in the rest of the figures~\ref{fig:figmatch} and
\ref{fig:fighowtomatch}, and there are still some gaps remaining after
different-$n$-complex CI is allowed\footnote{In
figure \ref{fig:figmatch} and \ref{fig:fighowtomatch}, the oscillation in
the background of the BPRM data can be eliminated with a larger number
of continuum basis functions \cite{z98}.}

\begin{table}
	\centering
	\caption {Selected levels of \fexvii ($J = 3, \pi = 1$) and \fexviii ($J=1/2, \pi=1$) to be matched (Note: the energy is $z$-scaled, and in unit of $10^{-2} Ry$)}
	\begin{tabular}{|c || c | c | c|}
		\hline
		& level index & BPRM & RDW \\
		\hline
\fexvii & 12 & -3.670657 & -3.67960 \\
		& 13 & -2.873930 & -2.84319 \\
		& 14 & -2.868775 & -2.84238 \\
		& 15 & -2.842915 & -2.83761 \\
		& 16 & -2.835625 & -2.83555 \\
		& 17 & -2.774853 & -2.77998 \\
		\hline
		\hline
		& level index & BPRM & RDW \\
		\hline
\fexviii & 64 & -1.275887 & -1.27703  \\
		& 65 & -1.234865 & -1.24175 \\
		& 66 & -1.226084 & -1.23724 \\
		& 67 & -1.136909 & -1.14691 \\
		\hline
	\end{tabular}	
	\label{tab:tablelevlsmatching}
\end{table}

\subsection{Bound-free data} \label{sec:sectionbf}
As BPRM calculations are carried out in the lower part of the whole energy 
range, they include low-$n\ell$ core configurations with $n \leq 4$. 
We use the RDW
method to extend it to higher regions up to 500 $Ry$ in photoelectron energy, 
and to include higher-$n\ell$ core configurations $n = 5,6$. The 
following part of this section gives detailed description of these aspects.

\subsubsection{High energy cross sections} \label{sectiontail}
\begin{figure}
        \centering
\includegraphics[width=\textwidth,height=9cm]{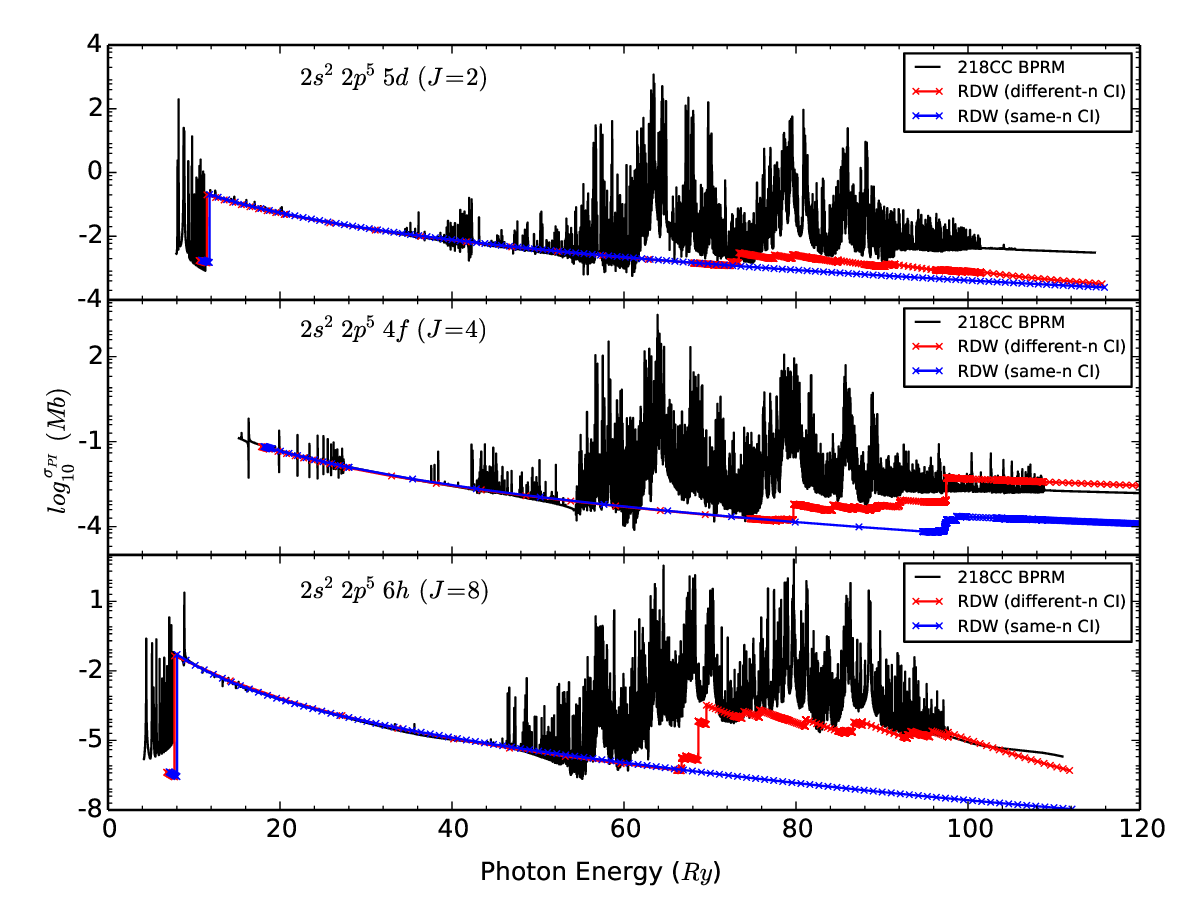}
\includegraphics[width=\textwidth,height=9cm]{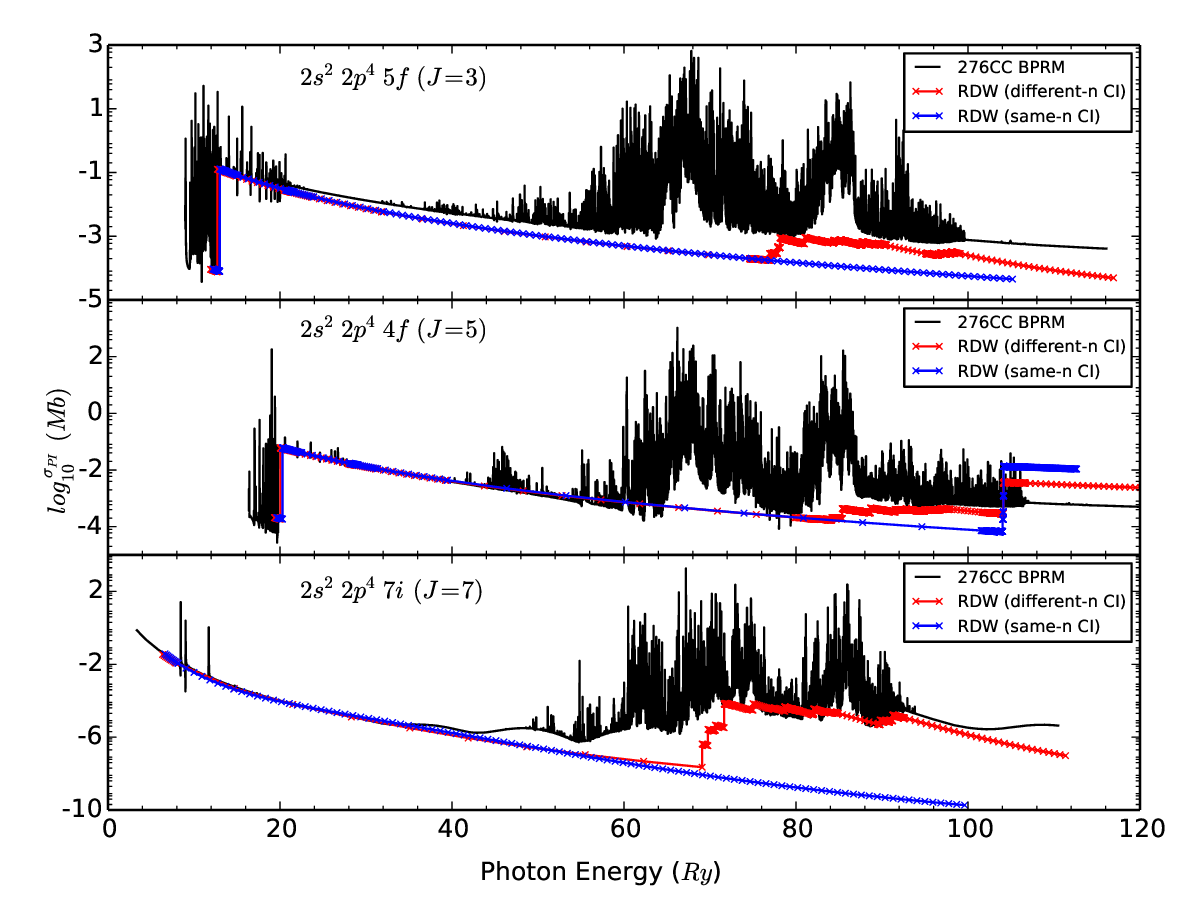}
        \caption{Different-n CI improves the
background significantly, but there is still very large discrepancy in
the right region of energy for some levels. BPRM (black), RDW(blue and
red). ``Same-$n$ CI'' refers to only same-n-complex configuration
interaction is considered for core configurations, and ``different-$n$
CI'' refers to both same-$n$- and different-$n$-complex configuration
interaction are considered.}
        \label{fig:fe17fe18discrepancy}
\end{figure}

\begin{figure}
        \centering
\includegraphics[width=0.9\textwidth]{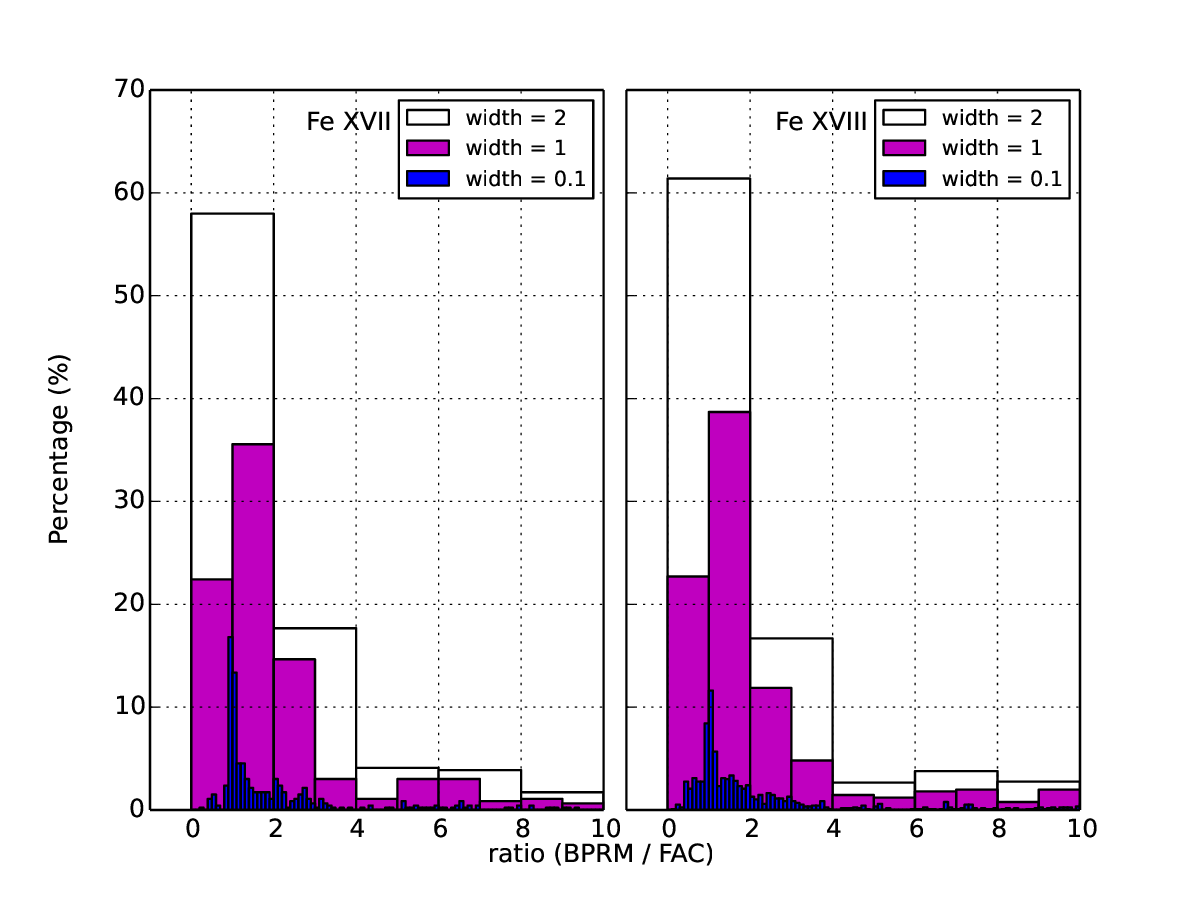}
        \caption{The distribution of the factors multiplied to RDW data
in the higher energy region for \fexvii and \fexviii.
``width'' is the width of the bins. }
        \label{fig:figratio}
\end{figure}

As shown in figure~\ref{fig:figmatch}, the RDW data can be matched almost
perfectly to background BPRM cross sections. However, we also find there
are cases where they don't match well in the right region of energy (see
figure~\ref{fig:fe17fe18discrepancy}). In the top panels
different-$n$
CI introduces many transitions, but they are not strong enough to raise
to the  background of BPRM. In the middle panel of 
figure~\ref{fig:fe17fe18discrepancy}, different-$n$ CI introduces many edges at
positions where the background of BPRM jumps, and raises the background
higher than BPRM. While in the middle panel of figure
\ref{fig:fe17fe18discrepancy}, around $105 Ry$, compared with same-$n$ CI,
different-n CI moves the background up on the left side, and down on the
right side, i.e. converging to the background of BPRM. As close-coupling
approximation treats CI more completely and accurately, we multiply the
RDW data in the higher energy region by a factor which is the ratio of
BPRM value and RDW value at the last point of BPRM calculation for each
level (see figure \ref{fig:figratio}), to account for the discrepancy
between BPRM and RDW. In figure \ref{fig:figratio}, the distribution of the
factors applied in the higher region is very alike for \fexvii
and \fexviii, and there are around $60\%$ of the bound levels
lying around ratio = 1. Among the high-ratio cases, some are caused by
the oscillation of the background of BPRM, which is due to the small
number of continuum basis functions used in the wavefunction expansion
\cite{z98} (see the bottom panels of figure
\ref{fig:fe17fe18discrepancy}. 

The energy mesh used in the region is created in such a way that 10 points are uniformly assigned between any adjacent ionization thresholds due to the other core configurations (see section \ref{sectionothertargets}).

\subsubsection{Highly excited core configurations} \label{sectionothertargets}
Using RDW with different-n-complex CI, we calculate photoionization
cross section due to other core configurations up to $n=6$ that are not
included in the BPRM calculation. To top-up 218CC BPRM for \fexvii, we
include core configurations $2s^2 2p^4 4f$, $2s 2p^5 4f$, $2p^6 4\ell'$,
$2s^S 2p^P 5\ell''$ and $2s^S 2p^P 6\ell'''$, where $\ell', \ell'',
\ell'''$ are all possible subshells in the corresponding shell, and $s$,
$p$ are any possible non-negative integers satisfying $S+P=6$. To top-up
276CC BPRM for \fexviii, we include core configurations $2p^5 3\ell'$, $2s^2 2p^3 4f$, $2s 2p^4 4\ell'$,  $2p^5 4\ell''$,  $2s^S 2p^P 5\ell'''$ and $2s^S 2p^P 6\ell''''$, where $\ell', \ell'', \ell'''$ and $\ell''''$ are all possible subshells in the corresponding shell, and $S$, $P$ are any possible non-negative integers satisfying $S+P=5$. 
The energy mesh is same as the one used in BPRM calculation 
merged with the one in the high energy region as described in section 
\ref{sec:sectiontail}. 

\begin{figure}
        \centering
\includegraphics[width=\textwidth]{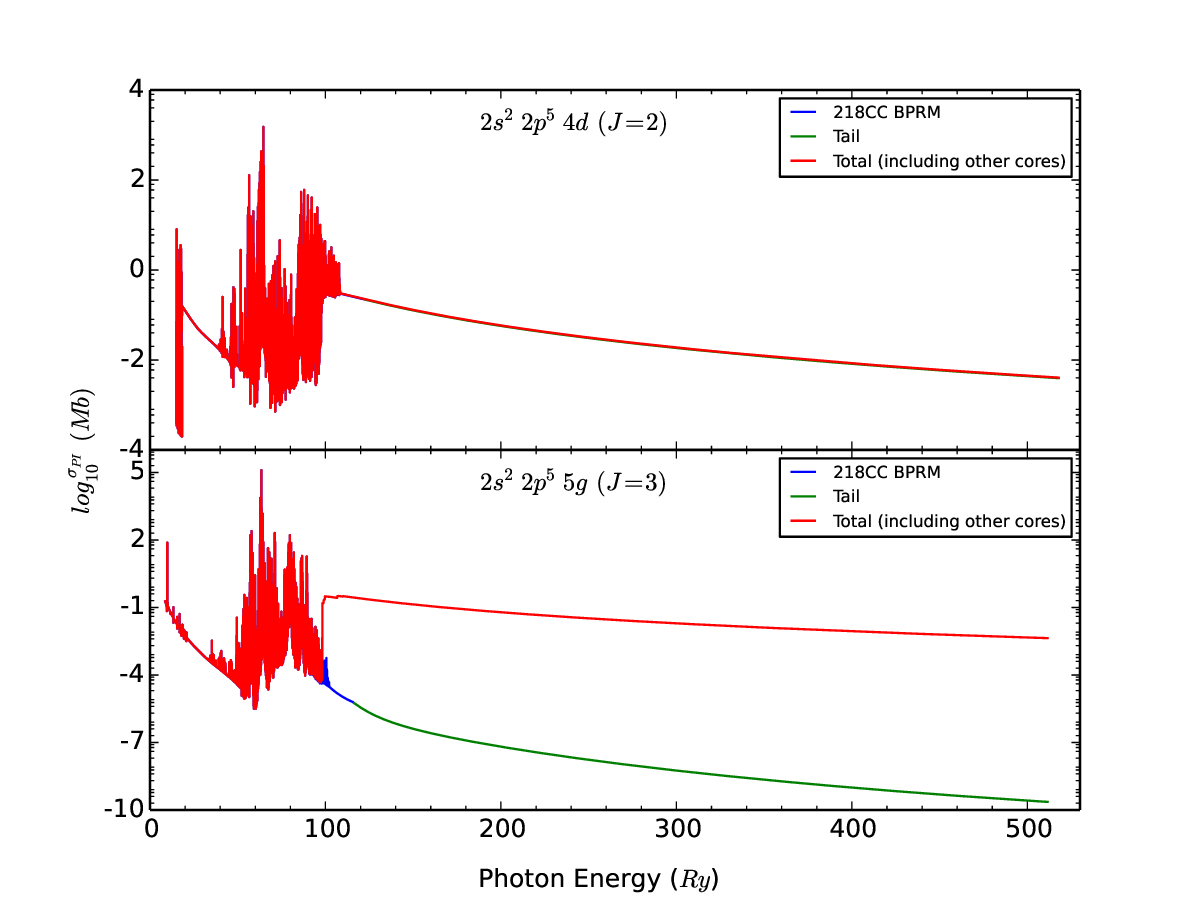}
\includegraphics[width=\textwidth]{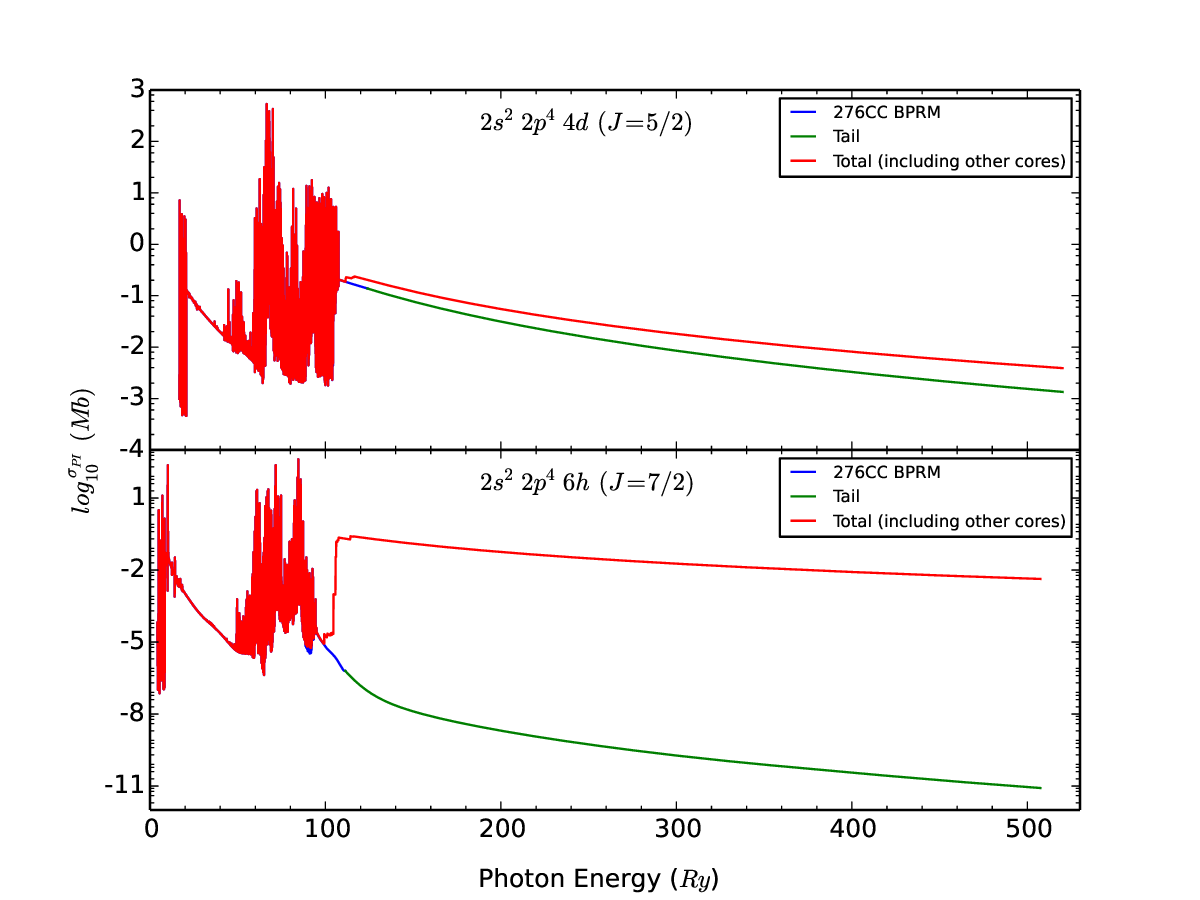}
        \caption{The photoionization cross section of the same four
levels as in figure \ref{fig:figmatch} are extended to higher energy region
and the contribution from other core configurations with
different-n-complex CI is added. Blue: BPRM calculation; green: RDW data
multiplied by a ratio; red: total photoionization cross section.}
        \label{fig:figfe17fe18tailother}
\end{figure}

As shown in figure \ref{fig:figfe17fe18tailother}, the BPRM data is merged with the scaled RDW tail, and the contribute from other core configurations varies from negligible to noticeable.


\subsubsection{Highly excited bound levels}
In RDW, we consider all the bound state levels with $n\leq10$, so we 
collect all such levels that are not included in BPRM calculation, 
and calculate the photoionization cross section due to all core configurations, 
i.e. the core configurations included in the BPRM calculation and the 
other ones displayed in section \ref{sectionothertargets}, with different-n-complex CI.

\subsection{Bound-bound data}
To top-up the bound-bound oscillator strength, we divide it into two
parts. One is from bound states to pure bound states. i.e. between negative
energy levels. We calculate all
such possible transitions, but only collect the ones that are not
calculated in BPRM calculations. The other part is from bound states to
quasi-bound states, i.e. from negative energy bound levels to positive
energy doubly excited states in the continuum. BPRM calculations treat
direct photoionization and autoionization as a single unified quantum-mechanical
process, as in section \ref{sec:sectionbf} that discusses direct
photoionization is done. To simulate autoionizing resonances, we
calculate the oscillator strengths from bound to doubly excited states,
among all pairs of negative-positive energies.
We consider transitions that excite an electron from $L$-shell to a
higher one, forming a doubly excited configuration that can not be
formed by combining a core configuration used in BPRM calculations 
with another electron. Take \fexviii for an example, we consider transitions 
from $2s^S 2p^P 3\ell$ to only $2p^5 3\ell' n\ell''$, where  $S$, $P$ are any 
possible non-negative integers satisfying $S+P=6$, 
and $\ell$, $\ell'$, $\ell''$ can be any sub-shell in the corresponding shell,
and $n= 3-6$. Since $2s^2 2p^3$ and $2s 2p^4$ are included in 276CC 
BPRM calculation, $2s^2 2p^3 3\ell' n\ell''$ and $2s 2p^4 3\ell' n\ell''$ 
are considered naturally. Thus they are excluded in the top-up
calculations.

 The number of quasibound positive energy levels in the continuum is far
larger than bound negative energy levels. For \fexvii we have 587 bound
levels as opposed to $\sim$72,000 positive energy levels included as
top-up. For \fexviii we obtain 1,154 bound levels vs. $\sim$175,000
quasibound levels. All possible oscillator strengths among these large
number of levels are computed and considered in opacities
calculations\footnote{We also note that oscillator strength data for
transitions among quasibound positive energy levels are employed
for free-free contribution to plasma broadening of autoionizing
resonances, discussed in paper RMOP3}.

\section{Conclusion}
In order to investigate the effect of convergence and completeness of
RMOP data for opacities calculations, complete sets of 
relativistic distorted wave calculations are carried out 
for \fexvii and \fexviii 
to compare with and topup the 218CC and 276CC BPRM calculations, respectively. 
Bound state levels are matched between BPRM and RDW calculation by comparing 
quantum numbers $J$, $\pi$, energy, and cross sections. 
With such level correspondence, BPRM photoionization cross sections
in the higher energy region are extended by scaled RDW data, 
and contribution from other core configurations up to $n=6$ is added, to
examine the effect of convergence over and above the $n \leq 4$ BPRM
data. 
Higher bound state levels are also included with photoionization cross
sections 
due to all core configurations up to $n=6$. Oscillator strength data
corresponding to the additional levels are also 
topped-up, with contribution from bound-bound and bound-quasibound transitions. 

The effects of CI on photoionization cross
sections are discussed, including same-$n$-complex and
different-$n$-complexes,
showing its significant role in reproducing the background of BPRM cross
sections using RDW. However, the extensive 
resonance structures that dominate BPRM photoionization cross sections 
throughout the energy range considered
can not be compared owing to their absence in the
RDW data. Nevertheless, the RDW method may provide useful checks on
completeness and convergence of CC-BPRM results.

 We have extensively studied the correspondence and complementarity
between BPRM
and RDW results with a view to ascertain possible impact on opacities.
However, the local-thermodynamic-equilibrium (LTE) 
Mihalas-Hummer-D\"{a}ppen equation-of-state 
valid in stellar interiors
yields extremely small occupation probabilities and level populations
for the high energy and high \eion spin-angular momenta states $nSLJ$
(discussed in paper P1), implying that the actual effect on opacities would
be small. Indeed, preliminary opacities calculations indicate that
Rosseland Mean Opacities are enhanced by only a
few percent $<$5\% (results to be reported elsewhere).

\section{Acknowledgments}
One of the authors (ZL) would like to thank Dr. Ming Feng Gu for helpful advice in using FAC and thank the following colleagues for helping run the BPRM calculation with their resources in ASC Unity cluster at the Ohio State University (names in alphabetic order): Jiaxin wu, Keng Yuan Meng, Max Westphal, Xiankun Li, Yonas Getachew and Zhefu yu. This research was supported by a teaching 
assistantship from the Dept. of Physics and the Dept. of Astronomy of Ohio State University (OSU) and by the US National Science Foundation and Dept. of Energy. Computations were carried out at the Ohio Supercomputer Center, the 
OSU Dept. of Astronomy and the ASC Unity cluster of OSU.

\end{document}